# A Gain Function for Architectural Decision-Making in Scientific Computing

Mariza Ferro, Antonio R. Mury, Bruno Schulze

National Laboratory of Scientific Computing
Getulio Vargas 333, Petropolis, Rio de Janeiro

Abstract

Scientific Computing typically requires large computational needs which have been addressed with High Performance Distributed Computing. It is essential to efficiently deploy a number of complex scientific applications, which have different characteristics, and so require distinct computational resources too. However, in many research laboratories, this high performance architecture is not dedicated. So, the architecture must be shared to execute a set of scientific applications, with so many different execution times and relative importance to research. Also, the high performance architectures have different characteristics and costs. When a new infrastructure has to be acquired to meet the needs of this scenario, the decision-making is hard and complex. In this work, we present a Gain Function as a model of an utility function, with which it is possible a decision-making with confidence. With the function is possible to evaluate the best architectural option taking into account aspects of applications and architectures, including the executions time, cost of architecture, the relative importance of each application and also the relative importance of performance and cost on the final evaluation. This paper presents the Gain Function, examples, and a real case showing their applicabilities.

Corresponding author

Email address: {mariza,aroberto,schulze}@lncc.br (Mariza Ferro, Antonio R. Mury, Bruno Schulze)

URL: www.lncc.br (Mariza Ferro, Antonio R. Mury, Bruno Schulze)

1. Introduction

Scientific computing involves the construction of mathematical models and numerical solutions techniques to solve complex scientific and engineering problems. The aim is to understand some natural phenomenon or to design a new device through simulations. Its importance is increasing and now is often mentioned as a third branch of science, complementing theory and experimentation [10].

Scientific computing generally requires a huge processing capacity in computer resources to perform large scale experiments and simulations with reasonable time. These large computational needs have been addressed with High-Performance Parallel and Distributed Computing (HPDC), which allows many scientific domains to leverage progress. There are many parallel architectures that can be used in order to achieve high performance, with many different designs, technologies and costs.

However, it is very di cult for many research groups to evaluate these HPDC architectures and get the best configuration to run their scientific applications. The reasons are:

> The different scientific domains have their scientific applications with different algorithms and mathematical models too, consequently, it requires different computational resources. So, it is a di cult task to de ne the best architecture to be acquired which can run a scientific application with the best performance. The task is especially hard when HPDC is not the expertise of the research group, because despite HPDC is essential to the advancing of scientific progress they do not want to specialize in that area.

> This task becomes even more complex when the acquired infrastructure should be used to a set of scientific applications. For many research groups in different countries, it is not possible to have a dedicated HPDC infrastructure to execute a scientific application. In this case, applications with its particular computational requirements, manipulating different input data (and so different execution times), are executed on the same computational architecture. Moreover, each application has a differentiated degree of importance for the development of the research. In this case, the researcher is confronted with a process of decision-making that may be need to be taken on the basis of multiple criteria.

Focused on this situation, when it is not possible to have a dedicated computational infrastructure, we developed a Gain Function (GF) that enables to measure and evaluate which is the best architecture to run a set of scientific applications. The GF can consider a number of applications, their performances (execution time), the relative importance to research and the acquisition costs. Furthermore, it is possible to consider the relative importance to the cost or performance on that choice.

The function was developed based on Utility Theory and on its concepts. It enables the assessment of various criteria and the evaluation of alternatives and then aggregation of these evaluations to achieve the relative ranking of the alter-natives with respect to the problem [22]. With its use, the problem of executing a set of scientific applications in the same infrastructure can be abstracted. It is possible to derive weights according to their impact on the research, and the objective of decisions to be made. Its applicability was evaluated in benchmark application examples and a in a real scenario where it illustrates its contribution to the decision-making.

The paper is organized as follows: In Section 2 the concept of Multi-Attribute Decision-Making and Multi-attribute Utility Theory approaches are presented. Also, the Analytic Hierarchy Process, used in our approach, is discussed and how to convert subjective assessments of relative importance to a set of overall scores or weights. Some related works are presented in Section 3. The details and mathematical proof of Gain Function are in Section 4, followed by experiments used to demonstrate its utilization. The Gain Function is applied in a real case for Bioinformatics domain and it is presented in Section 5. Finally, Section 6 concludes the paper and briefly discuss future work.

## 2. Background

In this section, we present the concepts and methods which are used as the basis to develop the Gain Function.

### 2.1. Multi-Attribute Decision-Making (MADM)

Decision-making can be defined as the choice, on some basis or criteria, of one alternative among a set of alternatives [22]. In this work, the decision-making problem is about what is the best computational architecture to execute a set of scientific applications. When the decision-making involves the decision about what is the best to execute a single application, the most important of all is the correct definition of the performance test, which one that actually represents the scientific application requirements [1]. In this case, the decision involves performance measures and architectural costs. But, when high-performance infrastructure is shared to execute a set of different applications the choice about the better architecture for that is a highly complex decision problem that may need to be taken on the basis of multiple criteria.

This is a multiple criteria problem since technology has created several alter-native architectures, with so many designs and costs. Additionally, the aspects of the scientific applications that can lead to much diverse performance results. Furthermore, when considering a set of users submitting a set of scientific applications, the importance of considering each research is relevant in which type of architecture must be acquired. In this case, decisions made in an ad-hoc manner have a high probability of being sub-optimal and so a formal decision-making methods is necessary.

The formal Multiple-Attribute Decision-Making (MADM) methods deal with decision-making in the presence of a number of often conflicting criteria and with a notion of alternatives and attributes. The alternatives represent different choices of action and attributes are referred as goals or decision criteria. These attributes may have totally different scales (qualitative or quantitative) and for each one is given a weight or relative importance with respect to their impact on the decision problem being solved [4].

---

[1] The important problem of defining a really representative execution test that represents the scientific application requirements is another problem where this Gain Function is inserted. However, this problem is not covered here, so it is considered that the proper test performance was executed.

The two main families in the MADM methods are those based on the Multi-attribute Utility Theory (MAUT). The MAUT methods consist of aggregating the different criteria into a function, which has to be maximized. The basis of MAUT is the use of utility functions that can be applied to transform the raw performance values of the alternatives against diverse criteria, both factual (objective, quantitative) and judgmental (subjective, qualitative), into a common, dimensionless scale. Utility functions can also convert the raw performance values so that a more preferred performance obtains a higher utility value. The Gain Function presented is this work was constructed based on utility functions concepts. For example, in our GF, when the criterion reflects the goal of cost minimization, the associated utility function must result in higher utility values for lower cost values [11] and higher performance.

In MAUT approaches, the weights associated with the criteria can properly reflect the relative importance of the criteria. In this work we use the Analytic Hierarchy Process (AHP), propose by [21], to convert subjective assessments of relative importance to a set of overall scores or weights.

The methodology of AHP is based on pairwise comparisons through questions that seek to define how important a criterion $C_i$ is relative to another criterion $C_j$. The answer to the questions of that following type are used to establish the weights according to the subjective (judgmental) criteria of the decision maker. The intensity of preference for one criterion versus another is given by the nine-point scale as presented in Table 1.

If the judgment is that criterion $C_j$ is more important than criterion $C_i$, then the reciprocal of the relevant index value is assigned. Let $c_{ij}$ denote the value obtained by comparing criterion $C_i$ relative to $C_j$ and as the criteria will always rank equally when compared to themselves, we have $c_{ij} = 1 = c_{ij}$ and $c_{ii}$ = 1. The entries $c_{ij}$, j = 1,..., m can be arranged in a pairwise comparison matrix C the size of mxm. Furthermore, with the weights determined by the pairwise comparison technique, the normalization of values is made and the relative preferences are obtained by the arithmetic average of each row of the matrix.

Table 1: Gradation scale for comparison of alternatives.

| Intensity of Preference | Numerical Scale |
|---|---|
| Extreme importance | 9 |
| Very strong importance | 7 |
| Strong or essential importance | 5 |
| Moderate importance | 3 |
| Equal importance | 1 |

In a very simple example with the comparison of only two criterion defined as $C_1$ = cost and $C_2$ = performance, the Table 1 can be used to derive the weights of the criteria and be arranged in a pairwise comparison matrix. If the judgment is that criterion $C_2$ it has a very strong preference in relation to criterion $C_1$, so $c_{11}$ = 1, $c_{12}$ = 1=7, $c_{21}$ = 7 and $c_{22}$ = 1. After making the normalization and the arithmetic average we obtain the numerical values of the criteria, $C_1$ = 0.125 and $C_2$ = 0. 875.

The AHP is a flexible decision-making process to help us to set priorities and make the best decision when both qualitative and quantitative aspects of a decision need to be considered. The concepts of Utility Functions and AHP were used to develop our Gain Function which is demonstrated next.

3. Related Work

There are some works that have some similarity of the ideia proposed in this work. Researches in market-oriented for grids and clouds, such as [27], [5] and [6], propose an analysis of the computational aspects based on qualitative aspects named as utility quantitative techniques. Consumers can specify their requirements and preferences for each respective job using Quality of Service (QoS) parameters and thus can assign value or utility to their job requests.

In this market-oriented approach indeed simply aiming to maximize utilization for service providers, and minimize waiting time for end users it captures the valuations that participants in these systems place on the successful execution of jobs and services. Therefore, the notion of maximizing the utility is also considered. This approach enables that specific needs of different users in order to allocate resources according to their needs. However, these works are only make a survey and proposed a taxonomy. In the work of [20] they propose a Petri net-based multi-criteria decision-making framework to assess a cloud service in comparison with a similar on premises service. The framework combines cost and qualitative issues to produce a final score and the aim is to employ a methodology simple for managers, to visualize and understand. These approaches are similar to ours in the sense of make an evaluation based not only on quantitative measures, but also in qualitative aspects (users preferences). However, the focus is the allocation of cloud and grids resources, while our approach is dedicated to high performance equipment selection.

Some works developed equations, named operational laws, to evaluate computer systems features like the classical Little's Law [7] and the more recent Processor Speed Up Law [23] and Occupancy Law [13]. The concept on which they are built involve only directly measurable quantities, like our mentioned operational values, and the concepts are from Operational Analysis, area that MAUT and Utility Theory are deeply inserted. The mathematical foundation is similar to ours Gain Function.

The work of [17], proposes a systematic and formal way of assessing the quality of an Enterprise Architectures (EA) based on decision maker's preferences. They propose that utility theory can be applied in EA meta models incorporating quality attributes. An explanation of how utility theory could be applied in EA models is provided and an example where two attributes (availability and cost) are evaluated. The authors make clear that it is a proper approach to supporting decision making with respect to different design alternatives for EA.

But, they use some ready equation from theory and a number of things are needed yet to make this approach functional. Also, it is different from ours in the application area.

The paper of [9] presents an approach to evaluation and selection of computer systems as a complex decision problem. Despite the approach don't use utility theory it is based on decision-making approaches and consider qualitative and quantitative features for decision. The mathematical models enables the use of subjective preferences by means of the aggregation function. The idea is similar to ours in the possibility of selecting the equipment that simultaneously satisfies buyer's cost and performance criteria. They consider individual hardware and software aspects for performance evaluation, such as disk memory capacity, main memory, network software and others detailed requirements. This characteristics is very useful by experienced professional but, as opposed to that, buyers of these systems are frequently less experienced prepared for this kind of system evaluation process. In this point this approach differs from ours, that enable not experience researchers in high performance architectures to make a decision.

So far it has not found any work that is most similar to ours, developing an evaluation function for decision-making on high-performance equipment.

4. Gain Function

As mentioned, in the reality of many research groups, a high-performance infrastructure shared to execute a set of applications and the choice about the best architecture for that is a highly complex decision problem that may need to be taken on the basis of multiple criteria. It is necessary to define the best architecture to be acquired, that meets the performance requirements of all applications, maximizing performance and minimizing costs.

The Gain Function developed in this work enables the decision-maker to determine which architecture delivers the maximum gain to execute a set of scientific applications with high confidence.

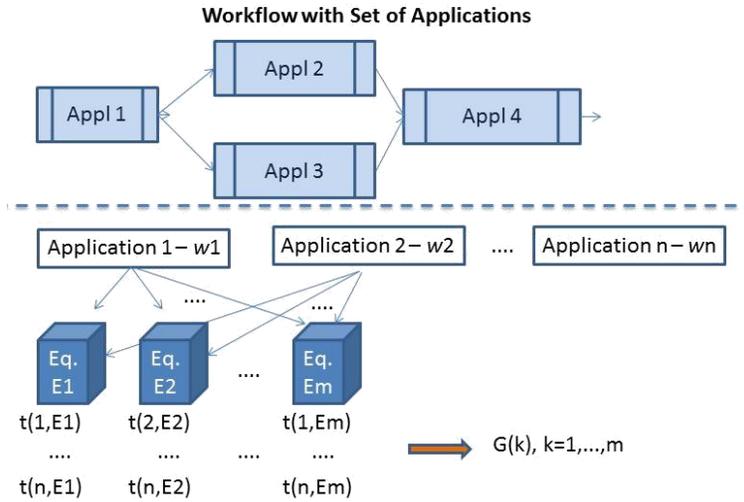

Figure 1: Example of a scenario where a set of applications (Application 1,…,n) and its respective weights ($w_1$,…, $w_n$) are executed to evaluate equipments/architectures (Eq. $E_1$,…,$E_m$).

Considering the problem of running a set of applications and for achieving that goal is considered the acquisition of new equipment/architecture. This scenario is exemplified in Figure 1, where the work ow of applications [2], that its execution is the objective, will be analyzed and for each application will be selected a representative performance test. Also, a weight for each application is assigned, which represents its relative importance on the work ow. The performance tests are executed in the sequence to evaluate each equipment and the operational values [3] (execution times) are collected. Further, it is possible to apply the Gain Function.

For each application j, j = 1,…,n, on each evaluated architecture $E_k$, k = 1,…,m, the execution time t(j; k) is measured. For each application j it is assigned a weight w_j. These weights are such $\sum_{j=1}^{n} w_j$ = 1. Also, for each that architecture is considered its cost c_k.

---

[2] It is important to note that when we mentioned work ow of applications we are not talking about a formal tool or model of scientific workflows. We are only talking about a sequence of applications that will be executed in certain order and it represents the applications used for researchers on shared form.

[3] Operational values are only directly measurable quantities, as opposed to e.g., probabilistic assumptions.

Let w_c and w_d be the weights for cost and performance, respectively, such that w_c + w_d = 1.

From those operational values, the GF enables to consider the performance (execution time) of each scientific application for each architecture evaluated. It is worth noticing that we are considering a relative importance for each application (applications' weights) and also the relative significance to the lower cost of the equipment or the best performance to run all applications. These preferences are de ned by the decision-maker by means of subjective assessments of relative importance about each application in relation to another one. The same is made about the relative importance of cost versus performance. Those subjective preferences are converted in a set of numerical weights by applying the AHP method, briefly discussed in Section 2.1.

Let $C_k = \frac{1}{c_k}$ be the relative cost for each equipment and $T(j,k) = \frac{1}{t(j,k)}$ be the execution time on each equipment k for each application j.

Note that, from these definitions, the equipment with higher cost (undesirable situation) has a small portion of the contribution in the function which represents the gain of each equipment. Similarly, the smallest execution time (desirable situation) has a large portion of the contribution on the final result of the Gain Function.

Next, $C_k$ and $T_{(j,k)}$ are normalized in order to make them dimensionless. We denote these normalized values by $C_{E_k}$ and $D(j, k)$, respectively. They are given by

$$C_{E_k} = \frac{C_k}{\sum_{p=1}^{m} C_p}, \; k = 1, \ldots, m,$$

$$D(j, k) = \frac{T(j,k)}{\sum_{p=1}^{m} T(j,p)}, \; j = 1, \ldots, n.$$

The GF is presented by the theorem below.

**Theorem 1.** *The Gain Function which represents the gain of each equipment $E_k$ when executing all the applications is given by*

$$G(k) = w_d \sum_{j=1}^{n} w_j D(j,k) + w_c C_{E_k}, k = 1, \ldots, m \qquad (1)$$

Proof: Lets define the gain function for an application j on an equipment k by

$$g(j,k) = w_c C_{E_k} + w_d D_{(j,k)}$$

When the weights $w_j$ are assigned for all the applications, we may extent this function in order to deal with all the applications and all the equipments, in the following way:

$$G(k) = \sum_{j=1}^{n} (w_j g(j,k))$$
$$= \sum_{j=1}^{n} w_j (w_c C_{E_k} + w_d D_{(j,k)}), k = 1, \ldots, m$$

G(k) represents the gain of the equipment $E_k$ when executing all the applications and their respective weights. It was also considered the weights for cost and performance.

Since $w_d$, $w_c$ and $C_{Ek}$ are independent of j, we can rewrite the Gain Function as:

$$G(k) = (w_c C_{E_k} \sum_{j=1}^{n} w_j) + (w_d \sum_{j=1}^{n} w_j D_{(j,k)})$$

As defined $\sum_{j=1}^{n} w_j = 1$ and the result is established. ∎

The equipment to be acquired is the one that presents the greatest gain, from Gain Function 1, i.e., the equipment $E_k$ such that

$$\max_{1 \leq k \leq m} G(k) = G(\bar{k}) \qquad (2)$$

Next, an example of how to apply the Gain Function is presented using data of experiments.

## 4.1. Example of Gain Function Application - Experiments

We conducted some experiments to verify the practicability of the Gain Function and the relevance of the results for decision making. Some experiments were conducted using two algorithms with distinct computational requirements on three parallel architectures to evaluate the GF applicability. Algorithms and architectures used in the experiments are described next.

### 4.1.1. Experimental Setup

The algorithms selected to evaluate the performance on parallel architectures were selected based on its computational requirements. For this objective we selected two classes of applications, based on Dwarfs characterization [18] that had been evaluated in our experiments to better understand the scientific applications [16] [15]. Dwarf classes represent applications with similar computational and data movement characteristics.

In this work, we intent to evaluate algorithms with distinct requirements, because this situation could be typical when a computational infrastructure is shared among a set of scientific applications. There are many scientific applications in many scientific areas classified in these classes [2]. We performed experiments using LUD (Dense Linear Algebra Dwarf class - computational intensive) and B+Tree (Graph Traversal Dwarf class - memory intensive) [3]. Those algorithms are available on Rodinia Benchmark suite [8] based on Berkeley's Dwarf. The default Rodinia's implementation was used for the tests, without any special setting up in the code for the multicore and manycore architectures. For this purpose the OpenCL [24] implementations were used, just for portability between the parallel architectures.

The experimental infrastructure used three architectures, summarized in Table 2 [4].

In each experiment presented, 30 runs were made for each point and the average and the standard deviation calculated. The confidence interval for the tests was less than 1%, so they are omitted in the results.

---

[4] We are not disclosing the commercial brands of architectures used because the objective of this work is not to evaluate and compare performance from different manufacturers.

Table 2: Target Architectures used in this work.

|  | x86 based Multi-core (Arch A) | x86 based Multi-core (Arch B) | Manycore GPU Based (Arch C) |
|---|---|---|---|
| Theoretical Peak Performance | 1177 GFlops | 281,6 GFlops | 1030 GFlops |
| Memory Bandwidth | \| | \| | 148GB/s |
| Memory Clock | \| | \| | 148 GB/s |
| Cores | 64 | 32 | 448 |
| Clock (GHz) | 2.3 | 2.2 | 1.15 |

Next, with the results of the experiments, we could evaluate a situation when two applications must share the same architecture. It is possible to de ne which is the best architecture, i.e., which one brings the major gain to execute these two applications.

We made the evaluations using the operational values (execution times and costs) and varying the relative importance between applications and between cost and performance.

These operational values are presented following the nomenclature used in the definition of the Gain Function 1. In Table 3, in the first column it is presented for each experiment (t(j, k)) conducted for each application j running on each architecture k (Table 2) followed by the average execution times on seconds (second column - Exec. Time (s)). The operational values for execution times corresponds to input data about the 18432x18432 matrix size for LUD (j = lud) algorithm and 30M (thirty thousand objects of data) for B+Tree (j = Btree). Despite the input data are so different, figuring out so different execution times, that is not a problem to be used in our GF, since these operational values are normalized (third column (D(j,k))).

Table 3: Average execution times and final normalized values.

| t(j, k) | Exec. Time (s) | D(j; k) |
|---|---|---|
| t(Btree, A) | 2489 | 0.15998 |
| t(Btree, B) | 813 | 0.48979 |
| t(Btree, C) | 1137 | 0.35022 |
| t(lud, A) | 347 | 0.25327 |
| t(lud, B) | 340 | 0.25714 |
| t(lud, C) | 180 | 0.48825 |

Based on the execution times presented in Table 3, it is possible to note that for the B+Tree algorithm the best performance was obtained when executing on architecture B. But, for LUD algorithm the best performance was obtained on architecture C. If the architecture that has been acquired is dedicated to one of these scientific applications, the test with the adequate benchmark and analysis of costs is enough to a decision-making. However, if the situation is to share the architecture of these two applications, the decision just looking for performances and costs is di cult. Additionally, if we want to consider the differences in relative importance between applications, the decision-making is even more difficult.

In these experiments, it is clear that architecture A seems to be the worst option, since its performance was the worst for all experiments and even with the highest cost. However, how to decide if architecture B or C offers higher gain is not so simple. Architecture B has the best performance for B+Tree application, but when executing LUD application its performance is worse than architecture C (almost twice as long). Furthermore, their costs are not quite different. In Table 4 are presented the costs for the acquisition of each architecture and its respective values $C_{Ek}$ to be used in Gain Function. In this table it is possible to see that the costs are very similar, making it di cult to decision making in this case [5].

Table 4: Operational values for costs and its values after normalization.

| c(k) | Value ($) | $E_k$ |
|---|---|---|
| c(A) | 8900 | 0.00011 |
| c(B) | 8760 | 0.00011 |
| c(C) | 8000 | 0.00011 |

A sensitivity analysis is presented in Table 5. In each column a set of weights for applications, cost and performance and the respective gain obtained for each architecture under these set of values. The weights presented in the first column represent a situation when there are no preferences between applications or between cost and performance. In this case, for all weights is assigned 0.5, i.e., they are equal. In the second column, we have an absolute preference for LUD application in relation to B+Tree and equal for cost and performance. These numerical values for preferences were assigned using the scale presented in Table 1 and next, to derive the weights, it was used a pairwise comparison matrix of AHP methodology (Section 2.1). In the other columns these weights are changed to evaluate different situations, and for all the conditions $\sum_{j=1}^{n} w_j = 1$ and $w_c + w_d = 1$ are complied.

The gain for each architecture was obtained for each weight variation through Function 1. So, in the decision-making we could consider the performances for the two applications, the cost of each architecture and relative preferences. The decision-making for this set of parameters pointed out that MAX(G(k)) (k = 1, 2, 3), in most of the situations, is for architecture C and only when B+Tree had absolute preference the gain pointed out for architecture B. In the first column, we can see that when the weights are equal the highest gain is offered by architecture C. The same occurs for the weights variations presented in columns 2,4 and 5. But, when B+Tree application presents an absolute preference in relation to LUD, the architecture B offers the best gain. This kind of result can have a significant impact on the decision about which equipment to be acquired.

---

[5] The costs were obtained from the suppliers to evaluate a real situation presented in next section.

Table 5: Sensitivity analysis: weights variation and respective gains for each architecture.

| Weights | | | | | |
|---|---|---|---|---|---|
| $w_{lud}$ | 0.5 | 0.9 | 0.1 | 0.5 | 0.5 |
| $w_{Btree}$ | 0.5 | 0.1 | 0.9 | 0.5 | 0.5 |
| $w_c$ | 0.5 | 0.5 | 0.5 | 0.7 | 0.3 |
| $w_d$ | 0.5 | 0.5 | 0.5 | 0.3 | 0.7 |
| Gain(k) | | | | | |
| Gain(A) | 0.26314 | 0.28179 | 0.24448 | 0.28574 | 0.24053 |
| Gain(B) | 0.34911 | 0.30258 | 0.39564 | 0.33937 | 0.35885 |
| Gain(C) | 0.38742 | 0.41502 | 0.35981 | 0.37469 | 0.40015 |

Beyond that, with the GF there are other possibilities. For example, it is possible to take the cost as an unknown variable and find, in front of the performances obtained, what is the maximum cost to match the maximum utility (gain).

In the next section the use of the Gain Function is exemplified in a real case of decision making about the best architecture for a set of Bioinformatics scientific applications.

## 5. A Gain Function Real Case - Experiments and Results

The objective of the experiments was to define a new infrastructure to execute a set of important applications for the Bioinformatics laboratory [6]. The new infrastructure will be shared between the applications, that have different degrees of importance to the researches. The Gain Function is applied to improve the decision-making about the best option and the experiments and results presented next.

---

[6] Labinfo - www.labinfo.lncc.br

5.1. Experimental Setup

The objective is the execution of three Bioinformatics applications, on acceptable execution time (time defined by researchers) to enable the development of a new pipeline of work, which needs more computational power and so a new computational infrastructure. The applications, the focus of the evaluation, were BLAST [1], MUMmer [14] and K-means [26].

BLAST (Basic Local Alignment Search Tool) is the main tool for compare a protein or DNA sequence to other sequences in various databases. It performs a pairwise sequence alignment between two protein or nucleotide sequences. BLAST searching allows the user to select one sequence (query) and perform pairwise sequence alignments between the query and an entire database (tar-get). In this search millions of alignments are analyzed and only the most closely related matches are returned. It is very important for researches in Bioinformatics and, in this case, it is fundamental to a whole-genome sequencing to study bacterial pathogens, which needs high computational performance. In the experiments performed in this work it was used a real input, in a FASTA format database, with thousands of sequencing reads against the NR database [7].

MUMmer is a software package that offers accurate alignments of entire genomes. MUMmer accepts two sequences as input and finds all subsequences that are no longer than a specified minimum length that are perfectly matched. The algorithm uses a suffix tree, which is a search structure that identifies all the maximal unique matches in the pairwise alignment [19]. The MUMmer is an important tool, but it is used for more specific analysis in the same project. In our performance experiments with MUMmer a FASTA format database was used as input, and it was executed between two different strains of Helicobacter Pylori with more than three millions of readings.

K-means is an unsupervisioning clustering algorithm extensively used in data mining. It is a partitioning method that constructs clusters based on a distance metric and is widely used for exploratory data analysis in Bioinformatics for structure finding in large databases. In the experiments performed in this work a database with more than nine million of data objects to be clustered was used as input .

---

[7] www.ncbi.nlm.nih.gov

According to constraints presented by the research laboratory, two multicore architectures were evaluated, summarized in Table 2. So, the applications were executed on architectures A and B and the execution times are presented in Table 6 and the costs were obtained from the suppliers (Table 4). The Gain Function was applied following the same reasoning presented in the example given in Section 4.1 and the architectures are evaluated.

Table 6: Average execution times for BLAST, K-means and MUMmer applications (j = blast; kmeans; mum).

| t(j; k) | Exec. Time (s) |
|---|---|
| t(blast, A) | 79341 |
| t(blast, B) | 193515 |
| t(kmeans, A) | 143 |
| t(kmeans, B) | 121 |
| t(mum, A) | 42 |
| t(mum, B) | 38 |

In both evaluations presented in Table 7, the gains were evaluated using equal weights for cost and performance ($w_c$ and $w_d$), because the researchers do not manifest preferences between them. In the first column, the weights between applications are equal too and the gains were evaluated without the preferences. In the second column the weights represent the different importance of the applications to research and it was defined in a verbal and subjective form by researchers. There is an absolute preference for BLAST application in relation to MUMmer and K-means and a moderate preference for MUMmer in relation to K-means. Then the numerical values of the weights were established using the AHP method, as presented in Section 2.

Table 7: Weights and respective gains for each architecture.

| Weights | | |
|---|---|---|
| $W_{blast}$ | 0.333 | 0.794 |
| $W_{kmeans}$ | 0.333 | 0.067 |
| $W_{mum}$ | 0.333 | 0.140 |
| $w_c$ | 0.5 | 0.5 |
| $w_d$ | 0.5 | 0.5 |
| Gain(k) | | |
| Gain(A) | 0.5190 | 0.5782 |
| Gain(B) | 0.4760 | 0.4223 |

If we look for the operational values (execution times – t (j, k) - Table 6) it is possible to note that BLAST application has a better performance on architecture A while MUMmer and K-means applications have the smallest execution time on architecture B. In this real situation, as performed on the examples, the results are divergent in terms of the best architecture when evaluating different scientific applications. As we saw in the examples, the costs are very similar and the decision making about what is the best architecture to fulfill the three applications requirements is difficult. So, the decision making was based on the quantitative values obtained with the execution time of the applications and architectural costs, added to the qualitative characteristics on the applications and the Gain Function 1 final results.

When the weights are equal, the major gain is presented by architecture A, despite the small difference. But, when the real relative importance assigned by researchers was considered the gain of architecture A is even greater.

However, this real case consider only three applications and two architectures, it is clear that the decision-making about that is not simple. If we con-sider only the acquisition costs, the selected architecture would be B. And if we only consider the performance of one unique application, as a traditional benchmark evaluation, the decision-making would be misguided too.

The researchers of Bioinformatics laboratory consider the use of the Gain Function very important to decision-making. They consider a paradigm shift in using this type of assessment, and not just considering theoretical peak performance values and costs associated with equipment. The Gain Function application was useful and it prevented that erroneous decision were taken.

It is important to consider all alternatives (aspects) involved in the decision about what is better for scientific research. The applications, their performances and the costs should be considered in a reliable decision about the computational architecture to be acquired.

6. Conclusions and Future Work

The decision-making about the best equipment of HPDC to attend a particular scientific research, one that really leverages the scientific progress is not a trivial task. Especially when this decision is not part of the research specialty. When this decision-making is about an equipment that will be dedicated to a scientific application, the decision process based on tests of performance value, and costs of hardware could be enough. However, this ideal situation, most of times, is not a reality. There are many research groups, in many countries, which most often, cannot acquire a dedicated equipment. It is shared for a set of scientific applications which could have contrasting computational requirements. In these situations the decision is much more complex, involving multiple criteria to be evaluated.

In this work, it was possible to note with the outcomes of the experiments, that applications have different performances when we change the architecture of the test. Moreover, the experiments show that when we considered the relative importance for application the decision could be totally changed. So, a reliable decision-making about the best one needs to consider a number of different architectures and applications, costs and subjective importance. In this case, it is not possible to make a decision based only on a simple performance and cost analysis. The use of a formal multi-attribute decision-making method is necessary to assist, both researchers and technicians, to obtain more reliable measurements.

In this work, we developed a Gain Function that assists the decision-maker to a decision with low risk and maximizing performance and minimizing costs. The function enables the evaluation of performance for a set of scientific applications on a set of equipments and also it is possible to consider the preferences between applications and also between cost and performance.

Thus, the use of formal methods such as the one presented here, developed under solid foundation of MAUT and AHP theory, has proved to be useful and relevant for scientific computing community. Decision-making based on the results evaluated by function allows to leverage scientific progress, because it is possible to determine the option that will really bring the best performance for researches, delivering this performance under the lowest cost.

In our future works the goal is to develop a new Gain Function that considers Green IT aspects [12] [25]. Besides the applications, performance and hardware cost, the function will also balance these criteria against power consumption. The gain measured when considering equipments and applications will also enable to maximize energy efficiency, besides the current performance maximization and cost minimization.

Acknowledgment

The authors would like to thank the financial support from the Coordination for the Improvement of Higher Education Personnel (CAPES), the National Counsel for Technological and Scientific Research (CNPq) and the Funding Agency of the State of Rio de Janeiro (FAPERJ) for the support of this project.